\documentclass[prd, aps, twocolumn, showpacs,superscriptaddress]{revtex4-1}

\usepackage{amsbsy,amsmath,amssymb,amsthm,wasysym,amsfonts}
\usepackage{graphicx}
\usepackage{psfrag}
\usepackage{epstopdf}
\usepackage{color}
\usepackage{mathrsfs}
\usepackage{comment}

\DeclareSymbolFontAlphabet{\mathrsfs}{rsfs}
\DeclareMathAlphabet\mathbfcal{OMS}{cmsy}{b}{n}

\newcommand{\be}{\begin{equation}}  
\newcommand{\ee}{\end{equation}}
\newcommand{\bea}{\begin{eqnarray}}           
\newcommand{\eea}{\end{eqnarray}} 
\newcommand{\beqn}{\begin{eqnarray*}}
\newcommand{\eeqn}{\end{eqnarray*}}
\newcommand{\ba}{\begin{align}}
\newcommand{\ea}{\end{align}}

\def\lm{{\ell m}}   

\def\E{{\cal E}}   
\def\J{{\cal J}}   
\def\F{{\cal F}}
\def\p4{{\psi_4}} 

\usepackage{float}
\definecolor{cyan}{rgb}{0,0.9,0.9}
\definecolor{orange}{rgb}{0.9,0.5,0}
\definecolor{magenta}{rgb}{1,0,1}
\definecolor{purple}{rgb}{0.8,0.4,0.8}

\begin{document}

\title{Energy versus Angular Momentum in Black Hole Binaries}

\author{Thibault \surname{Damour}}
\author{Alessandro \surname{Nagar}}
\affiliation{Institut des Hautes Etudes Scientifiques, 91440
  Bures-sur-Yvette, France}

\author{Denis \surname{Pollney}}
\affiliation{Departament de Fisica, Universitat de les Illes Balears, 
Palma de Mallorca, E-07122, Spain} 

\author{Christian \surname{Reisswig}}

\affiliation{Theoretical Astrophysics Including Relativity, California Institute of
  Technology, Pasadena, CA 91125, USA} 

\begin{abstract}
Using accurate numerical relativity simulations of (nonspinning) black-hole 
binaries with mass ratios $1:1$, $2:1$ and $3:1$ we compute the gauge invariant 
relation between the (reduced) binding energy $E$ and the (reduced) angular 
momentum $j$ of the system. We show that the relation $E(j)$ is an accurate 
diagnostic of the dynamics of a black-hole binary in a highly relativistic 
regime. By comparing the numerical-relativity $E^{\rm NR} (j)$ curve with the 
predictions of several analytic approximation schemes, we find that, while 
the usual, non-resummed post-Newtonian-expanded $E^{\rm PN} (j)$ relation 
exhibits large and growing deviations from $E^{\rm NR} (j)$, the prediction 
of the effective one-body formalism, based purely on known analytical 
results (without any calibration to numerical relativity), agrees strikingly 
well with the numerical-relativity results.
\end{abstract}
\date{\today}


\maketitle

{\it Introduction.} -- A ground-based network of interferometric gravitational 
wave (GW) detectors is currently being upgraded and is expected, thanks to an 
improved sensitivity, to detect, within a few years, the GW signals emitted 
during the inspiral and merger of compact binaries. The realization of this 
exciting observational prospect depends, however, on our theoretical ability 
to accurately compute, within Einstein's theory of general relativity, 
the motion of compact binaries and its associated GW emission. Recent 
developments have made it clear that the most efficient way to theoretically 
understand the late stages of the dynamics of compact binaries is to combine 
the knowledge coming from analytical relativity techniques, such as traditional 
post-Newtonian (PN) expansions~\cite{Damour:1983tz,Blanchet:1995ez,Damour:2001bu,Blanchet:2004ek}, 
or the newer effective-one-body (EOB) 
formalism~\cite{Buonanno:1998gg,Buonanno:2000ef,Damour:2000we,Damour:2001tu}, 
with the knowledge coming from numerical relativity (NR) simulations 
(see~\cite{Sperhake:2011xk} for a recent review). 
Here, we shall restrict our attention to binaries composed of 
two nonspinning black holes of masses $m_1$ and $m_2$. 
Our technique can, however, be applied to more general systems.

The aim of this Letter is to present how NR data can be used to explore, 
in a quite direct manner, the {\it dynamics} of black-hole binaries, by 
computing the relation between the total energy, $\E$, of the binary system, 
and its total angular momentum, $\J$. We compare the (gauge-invariant) 
relation $\E(\J)$ extracted from NR simulations to the corresponding 
analytical predictions from PN theory~\cite{Damour:1999cr}, 
and from EOB theory~\cite{Damour:2000we}.
We show that, during the inspiral (at least up to the last stable orbit) 
the gauge-invariant relation $\E(\J)$ is essentially independent of the 
current uncertainties in the analytic modelling of the emitted gravitational 
waveform, and can therefore inform us rather directly on the conservative 
dynamics of a black-hole binary. [This aspect of our work is akin to a 
recent study of periastron advance in black-hole binaries~\cite{Tiec:2011bk}.] 

{\it Numerical relativity.} -- Our results are based on new,
accurate numerical simulations of (nonspinning) black-hole binaries, 
which combine a $3+1$ Cauchy-evolved spacetime (using a variant of 
the ``BSSNOK'' evolution system, with moving punctures and an extended 
wave zone~\cite{Pollney:2009ut,Pollney:2009yz}) 
with a Cauchy-characteristic extraction 
(CCE) technique~\cite{Reisswig:2009us,Reisswig:2009rx}. 
The initial data for the $3+1$ evolution are conformally flat, 
Bowen-York Cauchy data, with the initial position and linear momenta
 of the punctures determined from a 3PN-accurate dynamical evolution 
starting from a large initial separation~\cite{Husa:2007rh}. 
These initial data lead to orbits having an eccentricity 
$e\sim 10^{-4}$. The CCE technique yields unambiguous 
estimates of the waveforms at infinity, without the need to 
extrapolate data extracted at finite radii. 
Here, we consider three simulations with mass ratios $q \equiv m_2 / m_1$ 
equal to 1, 2 and 3. The corresponding {\it initial} 
Arnowitt-Deser-Misner (ADM) total energy, $\E_0\equiv \E_{\rm ADM}$, 
total angular momentum, $\J_0\equiv \J_{\rm ADM}$ 
(oriented along the $z$ axis), and eccentricity 
are given in Table~I. 

We use these numerical simulations to compute the sequence of 
instantaneous values (at the retarded time $t$), $\E(t)$, $\J(t)$, 
of the system energy and angular momentum during the inspiral, 
by using the laws of conservation of $\E$ and $\J$ between the 
binary system and the emitted radiation. Namely, we compute
\be
\label{eq1}
\E^{\rm NR} (t) = \E_0 - \Delta \E_{\rm rad}^{\rm NR} (t) \, ,
\ee
\be 
\label{eq2}
\J^{\rm NR} (t) = \vert \mathbfcal{J}_0 - \Delta \mathbfcal{J}_{\rm rad}^{\rm NR} (t) \vert \, ,	
\ee
where the radiated energy and angular momentum, between the initial 
(retarded) time $t_0$ and time $t$, are computed from the multipole 
moments $N_{\lm}^{\rm NR}$ of the NR (complex) ``news function'' 
at infinity (we generally use units such that $G=c=1$):
\be
\label{eq3}
\Delta \E_{\rm rad}^{\rm NR} (t) = \frac{1}{16\pi} \sum_{\ell , m}^{\ell_{\rm max}} \int_{t_0}^t dt' \vert N_{\lm}^{\rm NR} (t') \vert^2 \, ,
\ee
\be
\label{eq4}
\Delta \J_{z \, {\rm rad}}^{\rm NR} (t) = \frac{1}{16\pi} \sum_{\ell , m}^{\ell_{\rm max}} \int_{t_0}^t dt' 
m\Im \left[ h_{\lm}^{\rm NR} (t') (N_{\lm}^{\rm NR} (t'))^* \right] \, .
\ee
Here $h_{\lm}^{\rm NR}$ is the NR multipolar metric 
waveform, $N_{\lm} (t) \equiv  dh_{\lm} (t) / dt$ and $\ell_{\rm max}=8$. 
We do not write here the expressions for the radiative losses of the other 
components $\J_x$, $\J_y$ of $\mathbfcal{J}$. We took them into account, though they 
turn out to have a negligible effect on the computation of $\J^{\rm NR} (t)$. 
While $\Delta \E_{\rm rad}^{\rm NR}$ only depends on the news function $N(t)$ 
(which is a direct output of the CCE code), the angular momentum loss also depends 
on the metric waveform $h(t)$. We computed (for each multipole) $h(t)$ from 
$\Psi_4(t)=dN/dt=d^2h/dt^2$ by the frequency-domain integration procedure 
of~\cite{Reisswig:2010di} 
(with a low-frequency cut-off $\omega_0 = 0.032 / (m_1 + m_2)$). 
In contrast to most studies of gravitational waveforms, 
we consider here the full time development of the GW emission 
from the start of the NR simulation, 
i.e., we crucially take into account the losses 
associated with the ``junk radiation'', 
viz the initial burst of radiation associated to the relaxation of the unphysical 
Bowen-York-type initial data, before the radiation settles 
down to a quasi-stationary inspiral pattern.

Finally, we replace the two $t$-parametrized series 
$\E^{\rm NR} (t)$, $\J^{\rm NR} (t)$ by the corresponding 
{\it unparametrized} curve $\E^{\rm NR} (\J)$. 
One example (for the mass ratio $q=1$) 
of our computations of the relation $\E (\J)$ is shown in Fig.~1. 
Here and below we work with the binding energy per reduced mass, 
$E \equiv (\E - M)/\mu$, and the dimensionless rescaled angular 
momentum $j \equiv \J / M\mu$, where $M \equiv m_1 + m_2$, $\mu \equiv m_1 m_2 / (m_1 + m_2)$. 
Fig.~1 compares the NR relation $E^{\rm NR}(j)$ to the predictions made by 
two different analytical formalisms: PN theory and EOB theory 
(as explained in detail below). 
The inset shows the very significant effect of the energy loss 
due to the junk radiation emitted at the beginning of the simulation. 
Note that $j$ decreases during the inspiral.
\begin{table}[t]
  \caption{\label{tab:tableI} Properties of the initial state of the NR simulations.}
\begin{center}
\begin{ruledtabular}
\begin{tabular}{ccccccc}
$q$  & $\nu$ & $e$ & $\E^{\rm NR}_0$ & $\J_0^{\rm NR}$ \\
\hline
$1$  & 0.25   & $1.5\times 10^{-4}$  & 0.9905197   & 0.9932560 \\ 
$2$  & 2/9    & $1.2\times 10^{-4}$  & 0.9908980   & 0.8559960 \\ 
$3$  & 0.1875 & $ 7.6\times 10^{-4}$ & 0.9933905   & 0.7675068 
\end{tabular}
\end{ruledtabular}
\end{center}
\end{table}

\begin{figure}[t]
\begin{center}
\includegraphics[width=0.4\textwidth]{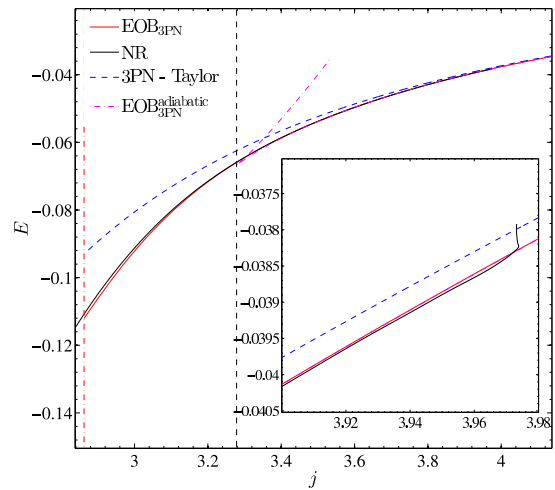}
\caption{\label{fig:fig1}Equal-mass case: comparison between 
four $E(j)$ curves. The standard ``Taylor'' PN curve shows the largest
deviation from NR results, especially at low $j$'s, while the
two (adiabatic and nonadiabatic) 3PN-accurate, 
non-NR-calibrated EOB curves agree remarkably 
well with the NR one.}
\end{center}
\end{figure}

{\it Post-Newtonian theory.} -- The gauge-invariant relation $E(j)$ 
has been computed in PN theory at the second post-Newtonian (2PN) 
approximation in~\cite{Damour:1988mr} and at the third post-Newtonian 
(3PN) one in~\cite{Damour:1999cr}. It has the structure
\be
\label{eq5}
E^{\rm PN} (j) = - \frac{1}{2j^2} \left[ 1 + \frac{c_1 (\nu)}{c^2 j^2} + \frac{c_2 (\nu)}{c^4 j^4} + \frac{c_3 (\nu)}{c^6 j^6} \right] \, ,
\ee
where $c_n (\nu)$ are polynomials (of order $n$) in the symmetric mass 
ratio $\nu \equiv \mu / M\equiv m_1 m_2 / (m_1 + m_2)^2$. 
[See Eq.~(5.1) of~\cite{Damour:1999cr}, completed by~\cite{Damour:2001bu} 
for the final determination of the 3PN dynamics, viz $\omega_{\rm static} = 0$.] 
The ``Taylor'' (i.e., nonresummed) $E^{\rm PN} (j)$ curve is shown in 
Fig.~\ref{fig:fig1} (for $q=1$) as a dashed line.

{\it Effective-one-body theory.} -- The EOB formalism maps 
the conservative dynamics of a two-body system onto the dynamics 
of one body of mass $\mu$ in a stationary and spherically 
symmetric ``effective'' metric, $ds_{\rm eff}^2 = -A (r;\nu) dt^2 + (A(r;\nu) \bar D (r;\nu))^{-1} dr^2 + r^2 (d\theta^2 + \sin^2 \theta d\varphi^2)$. The EOB potentials $A$ 
and $\bar D$ have been computed at the 2PN approximation 
in~\cite{Buonanno:1998gg}, and at the 3PN approximation 
in~\cite{Damour:2000we} (at 3PN one must complete the geodesic 
dynamics by terms, $Q(p)$, quartic in momenta). Here, we use 
the 3PN-accurate version of the EOB Hamiltonian, as defined 
in 2000~\cite{Damour:2000we}
(with $\omega_{\rm static} = 0$~\cite{Damour:2001bu}), 
i.e., with the effective-metric potentials $\bar{D}(u) \equiv 1+6\nu u^2 + (52 - 6\nu) \nu u^3$, 
and $A(u) \equiv P_3^1 \left[ 1-2u+ 2\nu u^3 + \left( \frac{94}{3} - \frac{41}{32} \pi^2 \right) \nu u^4 \right]$, 
where $u \equiv GM / (c^2 r)$, and where 
$P_3^1$ denotes constructing a $(1,3)$ 
Pad\'e approximant, so that $A(u)$ is a rational function of 
$u$ of the form $(1 + n_1 u) / (1+d_1 u + d_2 u^2 + d_3 u^3)$. 
In addition to the Hamiltonian dynamics defined by 
$A(u)$, $\bar D (u)$ (and $Q(u,p)$), the EOB formalism defines 
a radiation-reaction force $\F_{\varphi}$. Here, we use the ``newly resummed'' 
radiation reaction defined by~\cite{Damour:2008gu,Damour:2009kr}, with $3^{+2}$-PN accurate 
Taylor $\rho_{\lm}$'s, and {\it without} incorporating any ``next-to-quasi-circular''
(NQC) correction factor. The main point is that the resulting radiation-reaction-driven 
EOB dynamics uses only information that has long been analytically known, 
and does not rely on any information deduced from comparing EOB waveforms to NR waveforms. 
The resulting (nonadiabatic) 3PN-accurate, radiation-reaction driven EOB dynamics 
leads to the curve $E^{\rm EOB_{3PN}} (j)$ shown in Fig.~\ref{fig:fig1} as a 
red solid line. In addition, we also show the {\it adiabatic} EOB 
$E^{\rm EOB_{3PN}^{adiabatic}}(j)$ curve defined by considering the sequence 
of minima in $r$ (for a fixed $j$) of the (3PN-accurate) EOB Hamiltonian 
$H^{\rm EOB_{\rm 3PN}} (r,j)$. This adiabatic curve {\it only} depends 
on the potential $A(u)$ and has a cusp at the last stable orbit (LSO), $j_{\rm LSO}$.
The vertical distance $E^{\rm EOB_{3PN}} (j) - E^{\rm EOB_{3PN}^{adiabatic}} (j)$ essentially 
represents the kinetic energy linked to the (slow) inspiralling radial motion.
\begin{figure}[t]
\begin{center}
\includegraphics[width=0.4\textwidth]{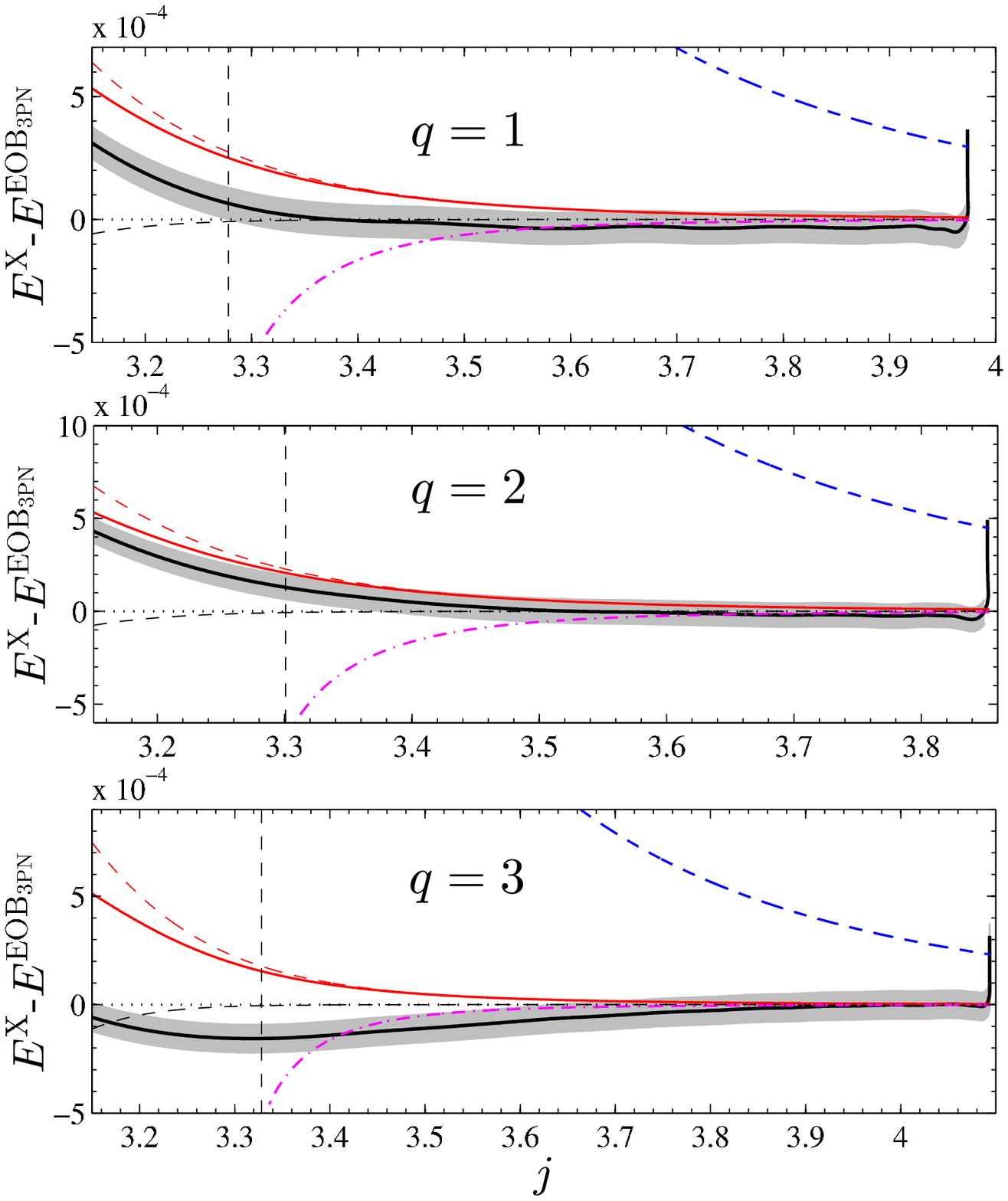}
\caption{\label{fig:fig2}Differences between seven $E^{\rm X}(j)$ 
curves and $E^{\rm EOB_{3PN}}(j)$, for the three mass ratios considered.
From top to bottom the labelling is:
${\rm X\,=\,PN}$, ${\rm EOB_{5PN}^{wo\,NQC}}$, ${\rm EOB_{5PN}^{NQC}}$, NR, 
${\rm EOB_{3PN}}$ (baseline), ${\rm EOB_{3PN}^{NQC}}$ and ${\rm EOB_{3PN}^{adiabatic}}$.
While the PN curve exhibits the largest deviations, all EOB curves remain
close to the NR one during the full inspiral, especially the 3PN-accurate,
{\it non-NR-calibrated} one.}
\end{center}
\end{figure}

{\it Results of the triple comparison NR-PN-EOB.} -- Fig.~\ref{fig:fig1} 
already exhibits several of the new results of our study: 
(i) The NR $E(j)$ curve starts at large $j$'s 
(i.e., large radial separations) close to the 
PN-predicted $E^{\rm PN}(j)$ curve, but then 
visibly deviates more and more from it during the inspiral [conventionally ending 
at the adiabatic(-EOB-defined) LSO, marked by a dashed vertical line]. 
(ii) By contrast the NR $E(j)$ curve is so close, on the scale of Fig.~\ref{fig:fig1}, 
to the (3PN-accurate, nonadiabatic) EOB prediction that their difference is barely visible 
not only during the inspiral, but also during the subsequent plunge. 
[The leftmost red vertical line in Fig.~\ref{fig:fig1} 
denotes the EOB ``light ring'', viz the end of the analytical 
inspiral-plus-plunge dynamics, and the beginning of the EOB description of 
the merger and ringdown.]
(iii) On the scale of Fig.~\ref{fig:fig1}, one cannot see, 
during the inspiral, the difference between the two EOB curves (nonadiabatic 
versus adiabatic). (iv) In addition, when zooming on the beginning of the 
$E^{\rm NR}(j)$ curve (see inset), we find that, although it coincidentally 
starts near the PN curve, it emits exactly the amount of junk radiation 
required to relax to the EOB prediction.
When considering the mass ratios $q=2$ and $q=3$, we obtained close analogs of 
Fig.~\ref{fig:fig1}, which exhibit exactly the same results (i)--(iv).

In order to refine and quantify these results, we henceforth zoom on the 
small deviations between the various $E(j)$ curves by using as horizontal 
baseline the (nonadiabatic, 3PN-accurate) EOB curve, i.e., by plotting the {\it differences} 
$E^{\rm X} (j) - E^{\rm EOB_{\rm 3PN}} (j)$, where the label X denotes either NR, PN, 
EOB$^{\rm 3PN}_{\rm adiabatic}$, or other EOB curves defined below. 
When focussing on the inspiral dynamics (above the LSO), this leads 
to NR-EOB differences of order $10^{-4}$, 
i.e., 300 times smaller than the $\simeq 3\times 10^{-2}$ change in the absolute 
value of $E$ during the inspiral, and 50 times smaller than the 
PN--NR difference $\sim 5 \times 10^{-3}$ at the LSO.
To discuss the meaning 
of the small NR-EOB differences, it is important to estimate the error attached 
to the NR $E^{\rm NR}(j)$ curve. We estimate an error on $E^{\rm NR}(j)$ by 
measuring the effect of changing, in turn, all the NR elements entering 
the computation of the losses Eqs.~\eqref{eq3}-\eqref{eq4}: (i) we replaced 
the CCE news by either the time integral of the curvature waveform 
$\Psi_4(t) = dN(t)/dt$ extracted at a large radius in the $3+1$ code, 
or a Regge-Wheeler-Zerilli metric waveform output by the latter code; 
(ii) we reduced the maximum multipolar order $\ell_{\rm max}$ used in the 
sums in Eqs.~\eqref{eq3}-\eqref{eq4} from the default value $\ell_{\rm max} = 8$ 
to $\ell_{\rm max} = 7$ and $\ell_{\rm max} = 6$; (iii) 
we varied the low-frequency cut-off $M\omega_0$ used in the 
frequency-domain computation of $h_{\lm} (t)$ from 
$\Psi_4^{\lm} (t)$~\cite{Reisswig:2010di} between 
about $0.01$ and $0.04$; (iv) we computed $h_\lm(t)$ from $N_\lm(t)$ instead
of $\Psi_4^{\lm} (t)$; (v) we explored the sensitivity to changes of 
the initial integration time $t_0$ in Eqs.~\eqref{eq3}-\eqref{eq4}; 
(vi) we replaced the high resolution NR data used as a baseline by 
medium resolution ones. 

Adding the effect of all these changes, and focussing on 
the crucial change in the energy loss $\Delta E_{\rm junk}$ 
linked to the initial burst of junk radiation, leads to a conservative error 
bar around $E^{\rm NR} (j)$ indicated by a gray-shaded region in Fig.~\ref{fig:fig2}.
In that figure, we plot the differences $E^{\rm X} (j) - E^{\rm EOB_{3PN}} (j)$ for 
$q = 1$, 2 or 3, and for six different labels X: NR (solid, thick, black curve), 
PN (upper, dashed blue curve), EOB$^{\rm adiabatic}_{\rm 3PN}$ (lower, dash-dotted magenta curve), 
EOB$^{\rm NQC}_{\rm 3PN}$ (black, dashed curve, just below the baseline), 
EOB$^{\rm NQC}_{\rm 5PN}$ (upper solid red curve) and EOB$^{\rm wo\, NQC}_{\rm 5PN}$ 
(dashed red curve, close to the previous one). Here, as above, the EOB baseline 
$E^{\rm EOB_{3PN}}$ (corresponding to the horizontal axis), 
as well as its adiabatic, EOB$^{\rm adiabatic}_{\rm 3PN}$, 
and NQC-completed, EOB$^{\rm NQC}_{\rm 3PN}$, avatars, use the 3PN-accurate 
EOB potentials of~\cite{Damour:2000we}. [EOB$^{\rm NQC}_{\rm 3PN}$ 
is defined according to the methods introduced in~\cite{Damour:2009kr} 
by adding a factor $f_{22}^{\rm NQC} (a_1 , a_2)$ in the $\ell=m=2$ mode, 
tuned to the maximum of the NR modulus.] 
Finally, $E^{\rm EOB_{5PN}^{NQC}}$ and $E^{\rm EOB_{5PN}^{wo\,NQC}}$
use the  NR-calibrated, 5PN potential 
$A_{\rm 5PN} (u) = P_5^1 [A_{\rm 3PN}^{\rm Taylor} (u) + \nu a_5 u^5 + \nu a_6 u^6]$, 
for $(a_5 , a_6) = (-6.3722, 50)$  [which lies in the middle of the ``good region'' 
of Ref.~\cite{Damour:2009kr}], either with (NQC) or without (wo NQC) NQC
corrections. Figure~\ref{fig:fig2} allows us 
to refine and strengthen the conclusions drawn above from Fig.~\ref{fig:fig1}, namely:
(i) The PN-predicted $E^{\rm PN}(j)$ curve disagrees strongly with the NR results; 
(ii) The 3PN-accurate nonadiabatic EOB curve, $E^{\rm EOB_{\rm 3PN}}$ 
(i.e., the horizontal baseline) 
is remarkably close to the NR results during the entire inspiral, 
with deviations that are smaller than the ``$2\sigma$'' level. 
(iii) The inclusion of nonadiabatic effects is 
important in continuing to ensure this agreement 
during the late inspiral
(see the difference $E^{\rm EOB^{adiabatic}_{3PN}} - E^{\rm EOB_{\rm 3PN}}$).
(iv) The inclusion of the NR-fitted NQC correction has a negligible effect 
during the inspiral: $E_{\rm wo\,NQC}-E_{\rm NQC}\lesssim 2\times 10^{-5}$.
(v) The EOB predictions based on the NR-calibrated, 5PN potential 
$A_{\rm 5PN} (u)$ of Ref.~\cite{Damour:2009kr} (with or without NQC corrections), 
are {\it less} close (especially for $q=1$ and 3) to the  NR 
result than the purely analytical 3PN-accurate EOB prediction. 
We verified that the same conclusion holds for the NR-calibrated 
5PN EOB potential suggested in~\cite{Pan:2011gk}.

{\it Summary.} -- We showed how to combine the knowledge of the initial (ADM) energy 
and angular momentum of a black-hole binary with accurate NR computations of its 
subsequent GW emission (including the initial burst of junk radiation), to derive 
the relation between the rescaled binding energy $E \equiv (\E - M) / \mu$ and the 
rescaled angular momentum $j = \J/(M\mu)$. Though the relation $E(j)$ does include 
nonadiabatic effects (linked to the radial kinetic energy during the inspiral, 
and thereby to the radiation reaction $\F_{\varphi}$) we have verified that the 
analytic uncertainties in the description of $\F_{\varphi}$ were essentially negligible 
during the inspiral, down to, at least, the LSO. This makes the NR-acquired knowledge 
of the $E(j)$ curve an accurate diagnostic of the {\it conservative dynamics} of a 
black-hole binary in a highly relativistic regime. By comparing $E^{\rm NR} (j)$ to 
various analytic descriptions of binary dynamics, we found that, while the usual, 
nonresummed 3PN-expanded relation $E^{\rm PN} (j)$ exhibits large and growing 
deviations with respect to $E^{\rm NR} (j)$, the EOB formalism, based purely on 
known analytical results (without NR calibration) predicts a relation $E^{\rm EOB} (j)$ 
which is remarkably close to $E^{\rm NR} (j)$. We found that the various 
existent {\it NR-calibrated} EOB formalisms fare somewhat less well than the 
purely analytic 3PN-accurate EOB formalism in agreeing with the NR results. 
This clearly shows that the NR curve $E^{\rm NR} (j)$ contains  
valuable information about the {\it conservative dynamics} of the binary during 
the inspiral that can usefully complement the information contained in the 
waveform (which mixes in an intimate manner dynamical and radiative effects). 
We leave to future work a detailed discussion of the information that can 
be extracted from $E^{\rm NR} (j)$.

{\it Acknowledgements} -- We thank Sascha Husa for assistance in
computing low eccentricity initial data. D.~P. was supported by grants 
CSD2007-00042 and FPA2010-16495 of the Spanish Ministry of Science. 
C.~R. was supported by the National Science Foundation under grant 
numbers AST-0855535 and OCI-0905046. Computations were performed 
on the NSF Teragrid (allocations TG-MCA02N014 and TG-PHY100033), 
the LONI network (\texttt{www.loni.org}) (allocation \texttt{loni\_numrel05}),
and the Caltech computer cluster ``Zwicky'' (NSF MRI award No.\ PHY-0960291).

\bibliography{refs}

\end{document}